\documentclass[aps,prd,twocolumn,superscriptaddress,footinbib,preprintnumbers,floatfix,showpacs]{revtex4}

\usepackage{dcolumn}
\usepackage{bm}

\usepackage{graphics}
\usepackage{graphicx}   
\usepackage{amsmath}
\input{epsf}

\newcommand\fverb{\setbox\fverbbox=\hbox\bgroup\verb}
\newcommand\fverbdo{\egroup\medskip\noindent%
			\fbox{\unhbox\fverbbox}\ }
\newcommand\fverbit{\egroup\item[\fbox{\unhbox\fverbbox}]}
\newbox\fverbbox

\newcommand{\be}{\begin{equation}}
\newcommand{\ee}{\end{equation}}
\newcommand{\ba}{\begin{eqnarray}}
\newcommand{\ea}{\end{eqnarray}}

\newcommand{\err}{\end{array}}
\newcommand{\bc}{\begin{center}}
\newcommand{\ec}{\end{center}}

\newcommand{\hmpc}{$h^{-1}${\rm Mpc}}

\newcommand{\eg}{e.g.,\,\,}
\newcommand{\ie}{i.e.,\,\,}

\newcommand{\lcdm}{$\Lambda$CDM }

\begin{document}
\preprint{}

 \vbox to 0pt{\vss
                    \hbox to 0pt{\hskip 440pt\rm ANL-HEP-PR-11-71\hss}
                  \vskip 45pt}

\title[Multi-stream flows]{The Cosmic Web, Multi-Stream Flows, and Tessellations} 
\author{Sergei Shandarin}
\affiliation{Department of Physics and Astronomy, University of
  Kansas, Lawrence, KS 66045}
\author{Salman Habib}
\author{ Katrin Heitmann}
\affiliation{High Energy Physics Division, Argonne National
  Laboratory, Lemont, IL 60439 }

\date{\today}

\begin{abstract}
  Understanding the structure of the matter distribution in the
  Universe due to the action of the gravitational instability -- the
  cosmic web -- is complicated by lack of direct analytic access to
  the nonlinear domain of structure formation. Here, we suggest and
  apply a novel tessellation method designed for cold dark matter
  (CDM) N-body cosmological simulations. The method is based on the
  fact that the initial CDM state can be described by a 3-D manifold
  (in a 6-D phase space) that remains continuous under evolution. Our
  technique uses the full phase space information and has no free
  parameters; it can be used to compute multi-stream and density
  fields, the main focus of this paper. Using a large-box $\Lambda$CDM
  simulation we carry out a variety of initial analyses with the
  technique. These include studying the correlation between
  multi-streaming and density, the identification of structures such
  as Zel'dovich pancakes and voids, and statistical measurements of
  quantities such as the volume fraction as a function of the number
  of streams -- where we find a remarkable scaling
  relation. Cosmological implications are briefly discussed.
\end{abstract}

\maketitle


\section{Introduction}
\label{sec:intro}

The large-scale structure of the universe observed in modern redshift
surveys like the 2dF Galaxy Redshift Survey and the Sloan Digital Sky
Survey displays a complicated geometry -- the `cosmic web' --
characterized by the presence of highly anisotropic concentrations of
galaxies known as filaments. Another conspicuous feature of the
large-scale structure is the presence of very large regions
practically devoid of bright galaxies that are conventionally called
`voids'. The filaments appear to be connected in a single percolating
network, often with clusters of galaxies positioned at the nodal
points. This complex structure is predicted by cosmological N-body
simulations, which yield a very similar distribution in both dark
matter halos that function as galactic hosts as well as in the dark
matter density field itself.

Because of the essential role of nonlinearity, cosmological N-body
simulations have become essential tools for studies of the formation
and properties of the cosmic web.  Collisionless gravity-only N-body
simulations can yield high-resolution information regarding the
distribution of matter and of dark matter-dominated halos. A first
principles approach to galaxy formation, however, requires a full
treatment of gas physics and processes such as star formation and
feedback from supernovae and AGN. Because of the difficulty of this
second step, analysis of observations is mostly performed by using a
phenomenological approach on top of gravity-only simulations. Thus
galaxies may be added to such simulations using halo occupation
distributions or semi-analytic models. Nevertheless, density fields
created from the particle or halo distributions can be useful for
morphological analyses of the formed large-scale structures (see \eg
Refs.~\citep{sch-wey-00,sh-she-sah-04,sch-07}).

Several different approaches exist to estimate density fields from a
particle distribution; many different methods have been suggested for
implementing these estimators. More than a dozen have been discussed
and compared in recent
publications~\citep{fer-etal-11,way-etal-11,mac-etal-09}. These
methods can be conveniently divided into two classes: those that
employ a tessellation of space \eg the Delaunay Tessellation Field
Estimator (DTFE) \citep{sch-wey-00,sch-07,pel-etal-03}, and those that
use particle-deposition in some form \eg Cloud-In-Cell (CIC)
\citep{hoc-eas-88} or the Smooth Particle Hydrodynamics (SPH) method
\citep{gin-mon-77,luc-77}. Here, we suggest a new approach based on
tessellation, that to our knowledge, has not been previously used in
cosmology.

Most density estimation methods use only the particle position
information; these are therefore limited to studies confined to
configuration space. However, incorporating particle velocities --
easily available in cosmological N-body simulations -- into the
analysis allows one to probe the full phase space, potentially greatly
advancing the understanding of the dynamics and morphology of the
cosmic web~\cite{ara-etal-04,asc-bin-05,mac-etal-09}.

Here we suggest the incorporation of dynamical information by using
the initial positions of the simulation particles together with the
final positions, in contrast to using the phase space information
available only at the final stage of the evolution. It is possible to
do this in CDM-based models of structure formation because the initial
condition is specified with a self-consistently determined vectorial
velocity field, uniquely defined at each spatial point.

In a classical Hamiltonian system, the initial coordinates and momenta
completely determine the path of the system in phase space. However,
it is well known that canonical transformations allow great freedom in
choosing generalized coordinates and momenta \citep{lan-lif-76}.  One
particular choice is to use the initial and final coordinates instead
of the initial coordinates and initial velocities, since both sets
completely define the path of the system in phase space. Therefore,
the set of the initial and final coordinates contains the same
dynamical information as the phase space information at the initial or
final time. Although this information is sufficient for the
determination of the phase space trajectory, it obviously has a very
different form. Therefore, the analysis must proceed along different
lines than those pursued in previous phase space-based studies
(Refs.~\cite{mac-etal-09} or \cite{ara-etal-04}).

It is important to emphasize that the (tracer) particles in a
cosmological N-body simulation have no intrinsic physical meaning. The
choice of their sizes and masses is dictated by computational
considerations alone. The physical equivalents of particles in N-body
simulations would be clouds of huge sizes, volumes, and numbers of
physical particles such as neutralinos or axions. These clouds are
certainly not rigid bodies nor uniform formations. In the real world, at the
initial stage of a simulation, they might be the structures
qualitatively similar to the final stage of the simulation but
significantly smaller in scale. It is well known that their overall
shape evolves: some get squeezed, others expand, and the shape
deformation is highly anisotropic, see \eg Ref.~\citep{kuh-mel-sh-96}.
Thus, using rigid particles in simulations is just a computationally
efficient technique to approximate the physical evolution of
collisionless self-gravitating systems.

The first major assumption in our approach is that the mass is not
concentrated in the form of a hard voxel around each point but is
relatively uniformly spread between the points. A typical particle
realization of an intially uniform mass distribution is to assign the
particles to the nodal points of a cubic grid (`quiet start'). In
simulations, the initial position and velocity perturbations are
implemented by applying the Zel'dovich approximation (or a
higher-order Lagrangian approximation) to this initial set of
points. In our case, the nodes of the cubic grid are interpreted as
the vertices of a set of tetrahedra that tile the space (at the
initial stage) without holes or other defects. This is the initial
tessellation of Lagrangian space.

The particles are then evolved using a standard N-body code. The
important difference compared to other approaches using tessellations
-- in most cases the Voronoi or Delaunay tessellations -- is that the
initial tessellation remains intact, preserving its initial neighbor
structure. Consequently, the 3-D hypersurface describing the initial
state of the system in 6-D phase space remains continuous and the
tetrahedra completely tile it at any stage of the evolution. In 3-D
configuration space, however, the tetrahedra can overlap
significantly, turn inside out, get squeezed, or expand without
limits. In order to estimate the density from the tessellation, we
have to make a second major assumption: the tetrahedra must remain
uniform.  Without knowledge of the structures on subgrid scales, this
seems to be the only sensible assumption. It sets the physical limit
to the maximum spatial resolution of the density and other fields.
 
Thus, in our approach we use the tessellation of Lagrangian (initial)
space and map it to Eulerian (final) space using the results of an
N-body simulation. As described below, our tessellation of Lagrangian
space is neither of the Voronoi nor of the Delaunay kind. It's one
point of similarity with Delaunay tessellation is that it tiles
Lagrangian space using tetrahedra with particles at the vertices. The
tessellation represents a continuous 3-D manifold in 6-D phase
space. Because of the symplectic nature of the underlying evolution,
the mapping to Eulerian space is continuous, and the manifold remains
three-dimensional and continuous despite the fact that it folds in
extremely complex ways during the nonlinear stages of the
evolution. The continuity of the manifold allows us to identify
multi-stream flows and to count streams, as well as estimate other
fields at an arbitrary point of Eulerian space.

The paper is organized as follows: we start with the description of
the tessellation we use and then explain how to compute the number of
streams and the density at an arbitrary point at a chosen instant of
time.  For illustrative purposes, we then compare the particle
distribution and the multi-stream and density fields in a small box
cut out from the full N-body simulation of a \lcdm cosmology. After this,
we present a statistical analysis of the tetrahedra parameters at the
nonlinear stage. We find the probability distribution function (pdf)
of the multi-stream field and density fields in regions with constant
number of streams and discuss their properties. Finally, we discuss
cosmological implications and summarize the results.

\begin{figure*}    
\includegraphics[width=17.5cm]{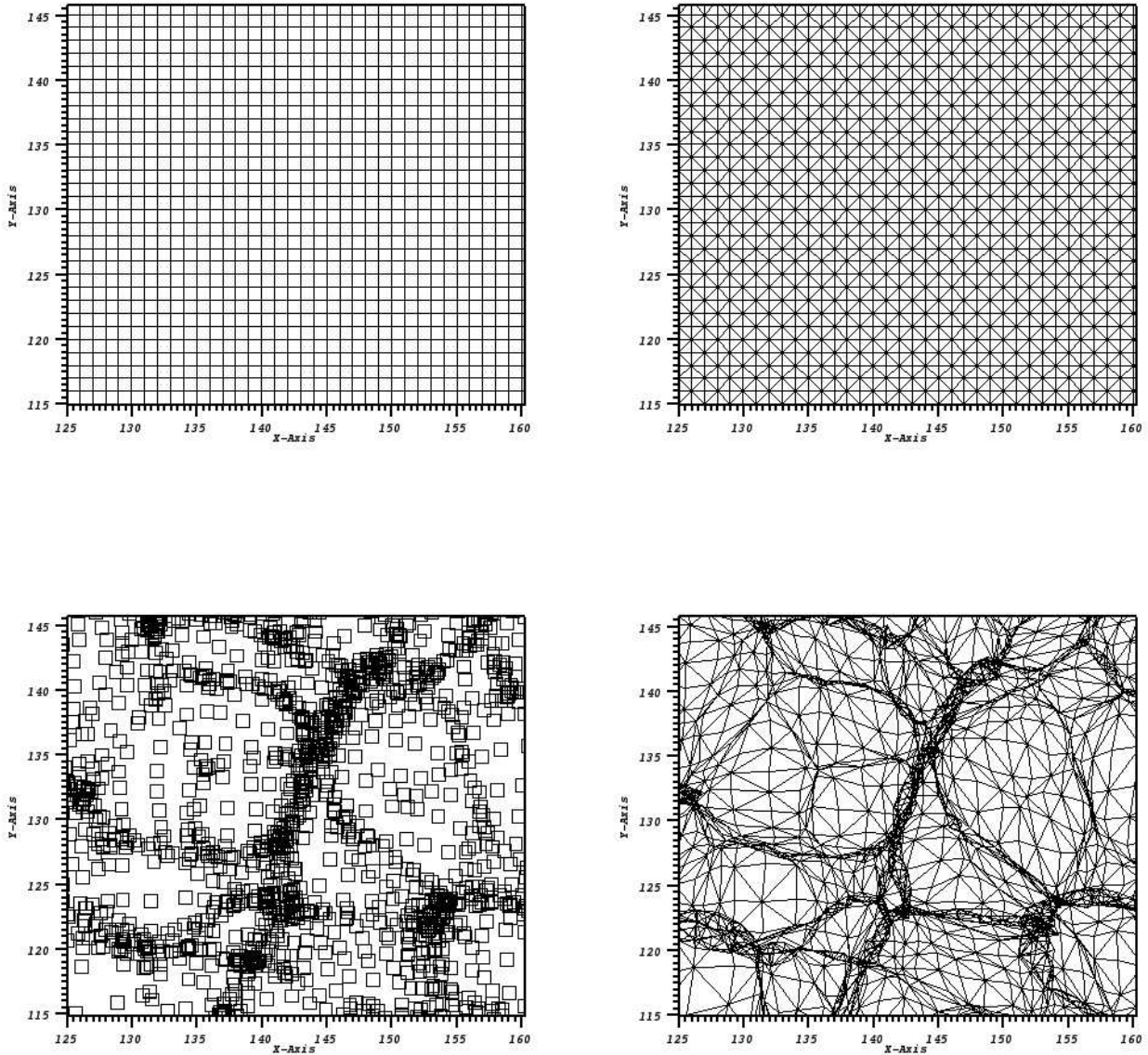}
\caption{Two-dimensional illustration of the difference in
  representing density fields by using particles (left column) and
  tessellations (right column). The top panels exhibit Lagrangian
  space where the density is uniform while the bottom panels display
  the system at a nonlinear stage of the  evolution. The nodes of the
  triangulation in the right panel are at the positions of the
  particles in the left panel.  The density field can be derived from
  the particle/node coordinates.}
\label{fig:meshes}
\end{figure*}

\section{Tessellations vs. particles}

The central aim of this work is to explore the prospects of using
multi-stream flows for better understanding the nonlinear regime of
the gravitational instability and the structure of the basic elements
of the cosmic web \ie clumps, filaments, pancakes, and voids, as
suggested in Ref.~\cite{sh-11}. We will describe a novel technique of
probing the multi-stream flows based on tessellation of the 3-D
manifold in 6-D phase space describing the evolution of a cold
collisionless medium -- significantly superior to the one used in
Ref.~\citep{sh-11}.

Figure~\ref{fig:meshes} provides a 2-D toy model illustration of the
main differences between using particles and tessellations.  We begin
with considering Lagrangian space, the common starting point for
generating initial perturbations in positions and velocities of
particles in cosmological N-body simulations
\citep{dor-etal-73,dor-etal-80,kly-sh-83}.  The homogeneous state is
approximated by generating a regular square mesh with particles placed
at the nodes as shown in the top left panel. The ensuing evolution of
the system is completely described by the motion of the particles.
The particles can be assigned sizes as illustrated by a CIC example in
the bottom left panel.

Alternatively, the nodes of the initial mesh can be considered as the
vertices of a triangular tessellation as shown in the top right panel
of Fig.~\ref{fig:meshes}. The motion of the particles/nodes results in
the displacement of the triangles along with their deformation (bottom
right panel).  The bottom panels demonstrate the similarities as well
as the main differences of the two approaches. The generic
disadvantage of the CIC representation is related to two
problems. First, structures smaller than the `particle size' are
filtered, and entirely empty regions can be produced that are
unphysical in the context of dark matter dominated cosmological
models. Adaptive gridding schemes can help alleviate these problems
but do not completely eliminate them. In addition, they typically
involve one or more free parameters. For our purposes, a tessellation
approach is therefore more natural.

\begin{figure}    
\includegraphics[scale=0.5]{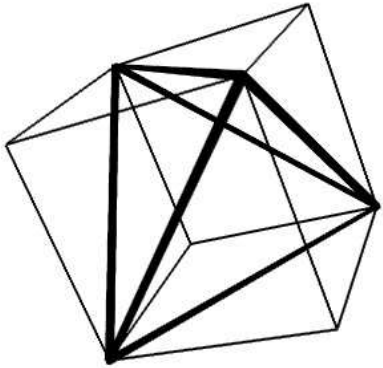}
\caption{Decomposition of a cube into five tetrahedra. The central
  tetrahedron shown by a heavy line is regular and its volume equals
  a third of the volume of the cube. Each of the remaining four
  tetrahedra is half of the volume of the central one, or one-sixth of
  the volume of the cube. }
\label{fig:cube}
\end{figure}

Now we turn to the case of 3-D space. As in the case of two
dimensions, we start with a cubic mesh in Lagrangian space and place
particles at the nodes of the mesh. Then we decompose each elementary
cube into several tetrahedra as shown in Fig.~\ref{fig:cube}.  The
goal is to tessellate a continuous 3-D manifold or hypersurface in a
6-D phase space without holes or other defects.  This requires that
the faces and edges of the adjacent tetrahedra must match to each
other. In order to achieve this goal, the adjacent cubes that have a
common face must be rotated by 90 degrees with respect to each
other. It is worth stressing that the tessellation employs only
particles \ie the nodes of the Lagrangian tessellation.  This
guarantees that during further manipulations we will use only
information contained in the positions and velocities of the
particles.  Out of several possible options we have chosen to
decompose the fundamental cube into five tetrahedra as shown in
Fig.~\ref{fig:cube}.  A minor disadvantage of this choice is the
presence of two kinds of tetrahedra differing in their shapes and
volumes.

The advantages of using tessellations -- in particular Voronoi and
Delaunay tessellations -- for studying the cosmic web have been
investigated and described in detail by van de Weygaert and
collaborators (see \eg
Refs.~\cite{ara-cal-etal-07,pla-etal-11,sch-wey-00,sch-07} and
references therein). We repeat that the tessellation we
use here is neither of the Voronoi nor Delaunay type. The only analogy
of our tessellation with the Delaunay tessellation is that both are
3-D triangulations -- as all elements of the tessellation are
tetrahedra.  We also wish to stress another key difference: we build
our tessellation only once in Lagrangian space. In the dynamical
evolution that follows, the same sets of nodes define each tetrahedron
as in Lagrangian space.  As a result, the neighboring tetrahedra
always share the same faces, edges, and vertices. The tetrahedra may
arbitrarily change their sizes, be turned inside out many times and
overlap in the course of the evolution of the system, but the initial
3-D manifold remains continuous. Preserving continuity of the 3-D
manifold allows us to identify multi-stream flows and easily count
streams at arbitrary points in Eulerian space.

\section{Computing multi-stream and density fields from the tessellation}
\label{sec:multi-stream_field}

The nonlinear stage of gravitational instability in a collisionless
CDM model is characterized by the emergence of regions where more than
one stream (or flow) can coexist at the same point in space.
Zel'dovich~\citep{z-70} was the first to discuss this phenomenon in
the context of the formation of structure in the Universe. In
particular, Zel'dovich predicted the formation of very oblate
concentrations of mass which he baptized as `pancakes'. The pancakes
are the first unambiguously nonlinear structures to grow from generic
initial density fluctuations in a continuous cold medium under the
action of gravity, see e.g., Ref.~\cite{sh-etal-95}.


\subsection{Multi-stream field}

We define a multi-stream field as a field taking discrete values that
are equal to the number of streams at every evaluation point. At a
generic point the number of streams is always an odd integer. An even
number of streams occurs only on caustic surfaces \ie on a set of zero
volume. As the nonlinear stage progresses, the three-dimensional
manifold becomes more wrapped and folded in phase space, especially in
the regions of high density such as dark matter halos, where the
dynamical timescale is short.  When a new fold emerges, the number of
streams increases by two at every point in the region of the newly
formed fold.  The formation of new streams indicates the advance of
nonlinearity and therefore the multi-stream field is an objective
quantitative characteristic of nonlinear evolution \cite{sh-11}.

A typical local characteristic of nonlinearity in cosmological studies
is simply the density. The root mean square deviation of density,
$\sigma_{\delta} = \langle \delta^2\rangle^{1/2}$, where $\delta
\equiv (\rho-\bar{\rho})/\bar{\rho}$, is commonly used as a global
characteristic of nonlinearity. While the virial overdensity factor
value of $\sim$200 is grounded in the dynamics of halos, in general
the value of $\sigma_{\delta}$ has a rather vague relation to the
complex dynamics that occurs between the linear stage and the virial
equilibration of halos.  It appears that the multi-stream field can
describe some aspects of nonlinearity differently -- and sometimes
better -- than the density, especially in pancakes and filaments
\cite{sh-11}.

If the tessellation of the volume occupied by the system is
established in the homogeneous state or the linear regime, 
then one can compute the multi-stream field at any time from the
current coordinates of the particles without following their
trajectories. Let us consider an arbitrary point in Eulerian
space. The number of streams at this point is equal to the total
number of tetrahedra for which the chosen point is internal. Whether a
chosen point is inside or outside a tetrahedron can be established by
computing five determinants as described below. 

If the coordinates of the chosen point  are $(x,y,z)$, and the
vertices of a certain tetrahedron have the  coordinates
$(x_i,y_i,z_i)$ ($i=1,2,3,4$),  then the determinant computed using
the coordinates of the vertices 
\be
d_0=\begin{vmatrix}
x_1 & y_1&  z_1& 1 \\
x_2 & y_2&  z_2& 1 \\
x_3 & y_3&  z_3& 1 \\
x_4 & y_4&  z_4& 1  \end{vmatrix}~~~
\label{eq:d0}
\ee
and four determinants that also use the coordinates of the point
itself 
\ba
d_1=
\begin{vmatrix}
x& y&  z& 1 \\
x_2 & y_2&  z_2& 1 \\
x_3 & y_3&  z_3& 1 \\
x_4 & y_4&  z_4& 1  \end{vmatrix},~~~
d_2=
\begin{vmatrix}
x_1 & y_1&  z_1& 1 \\
x& y&  z& 1 \\
x_3 & y_3&  z_3& 1 \\
x_4 & y_4&  z_4& 1  \end{vmatrix},~~~ \nonumber\\
d_3=
\begin{vmatrix}
x_1 & y_1&  z_1& 1 \\
x_2 & y_2&  z_2& 1 \\
x& y&  z& 1 \\
x_4 & y_4&  z_4& 1  \end{vmatrix},~~~ 
d_4=
\begin{vmatrix}
x_1 & y_1&  z_1& 1 \\
x_2 & y_2&  z_2& 1 \\
x_3 & y_3&  z_3& 1 \\
x& y&  z& 1 \\  \end{vmatrix}~~~~ 
\ea
all have the same sign, then the point is inside the tetrahedron. The
determinant $d_0$ also gives the volume of the tetrahedron  $V =
|d_0|/6 $. If $d_0=0$, then four vertices of the tetrahedron are
coplanar and its volume  is zero. If $d_0<0$, then it has turned
inside out an odd number of times. If any other $d_i=0$ ($i=1,2,3,4$),
then the  point lies on the face $i$  which is opposite  to the vertex
$i$. If the sign of $d_i$ is different from that of  $d_0$, then the point
lies outside the boundary of face $i$, otherwise it is
inside. Formally the multi-stream field can be computed with arbitrary
resolution, however, the real physical accuracy of such a computation
is ultimately determined by the mass and force resolution of the
simulation. 
 
\subsection{Density field}

The method for computing the multi-stream field can be easily extended
for computing the density. After finding that a point resides inside a
particular tetrahedron, one can compute the volume of the tetrahedron,
and therefore the associated density, assuming that the mass within
every tetrahedron is conserved and its density is uniformly
distributed.  Summing up the densities over all streams one finds the
density at the chosen point. As in the case of the multi-stream field
the density field can be computed on an arbitrary mesh or on an
arbitrary irregular set of points.
 
\section{Multi-stream field in $\Lambda$CDM}

We now apply the technique described above to study aspects of the
large-scale structure in a standard $\Lambda$CDM cosmology. Because
our main aim here is to introduce the technique, we restrict ourselves
to a relatively small number of illustrative examples.

\subsection{The cosmological model and simulation parameters}
The formation of structure is modeled here by a gravity-only N-body
simulation in a 512\hmpc-sided cubic box using the HACC (Hardware
Accelerated Cosmology Code) framework~\cite{hab-etal-09,pop-etal-10}
in its medium-resolution, particle-mesh (PM) mode. The number of
particles is 512$^3$ and the grid size in the gravitational Poisson
solver is 1024$^3$. The parameters of the $\Lambda$CDM cosmological
model are as follows: $h = H_0/(100 \, {\rm km/s}\cdot {\rm Mpc}) = 0.72$,
$\Omega_{tot} = 0.25$, $\Omega_b= 0.043$, $n = 0.97$, $\sigma_8 =
0.8$, the initial redshift $z_{\rm in} = 200$. These parameter values
are roughly consistent with current measurements; our aim here is not
to test the standard cosmological model but to provide a more or less
realistic environment within which to try out the new technique. For
ease of analysis, in the following we will restrict ourselves to a
relatively modest spatial dynamic range (hence the choice of the HACC
PM mode made here). Analysis of high-resolution HACC simulations will
be reserved for a later publication.

\subsection{Small box study}
We first demonstrate our new method by studying a small cubic box cut
out from the main simulation volume described above. The size of the
small box is 8$h^{-1}$Mpc. We shall see later that there are
several pancakes in this box and that they form a quite complicated
structure; even at this relatively small scale, they overlap in
projection, obstructing the structures behind them. Because of this
problem, structure in a box of significantly greater size becomes
difficult to visualize. Hence, it is useful to first use a small box to
illustrate the properties of our method.

The mean density in the chosen box is $\rho_{\rm box} = 1.78
\bar{\rho}$ where $\bar{\rho}$ is the mean density in the
Universe. The total number of non-overlapping boxes of this size in
the simulation volume is $64^3=262144$; this allows us to obtain quite
accurate statistics describing the regions of this shape and size. The
Universe is highly inhomogeneous on the box scale. The mass, and
therefore mean density, $\rho_{\rm box}$, of the boxes have a highly
nongaussian pdf as can be seen from the following numbers: ${\rm
  min}(\rho_{\rm box}) =0.043\bar{\rho}$, ${\rm median}(\rho_{\rm
  box}) = 0.62\bar{\rho}$, $\langle\rho_{\rm box}\rangle =
\bar{\rho}$, ${\rm std}(\rho_{\rm box}) = 1.32\bar{\rho}$, and ${\rm
  max}(\rho_{\rm box}) = 61.7\bar{\rho}$. The chosen box contains
roughly 80\% more particles the the average box with the same size
(i.e., it is roughly 80\%  more massive).  About 13\% of such boxes,
i.e., slightly more than 34,000 boxes overall, have greater masses.

\begin{figure}
\includegraphics[width=10cm]{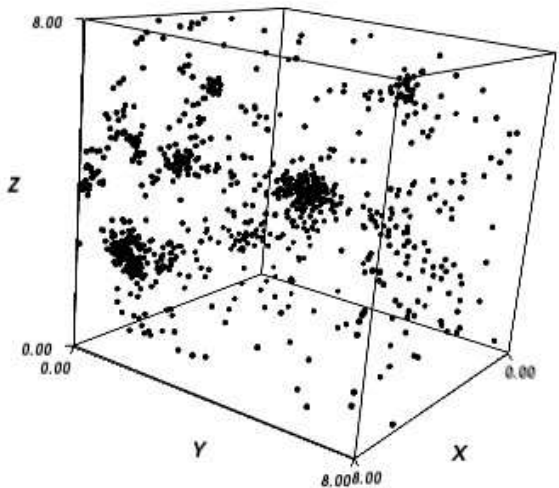}
 \caption{Particle distribution \ie tetrahedron vertices  in 
 a small sub-cube of size 8$h^{-1}$Mpc  extracted from the full
 simulation.} 
\label{fig:particles}
\end{figure}
\begin{figure}
\includegraphics[width=10cm]{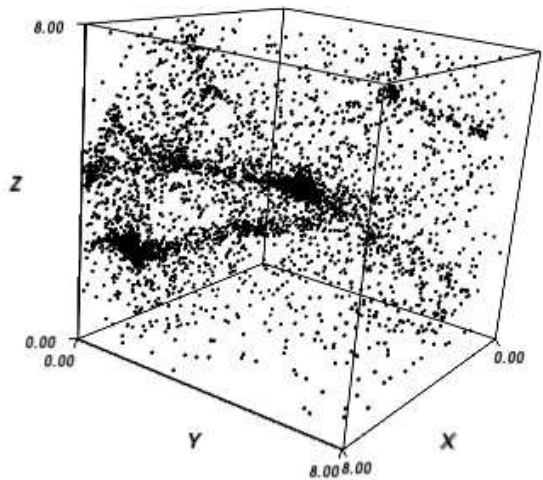}
 \caption{
  Distribution of the tetrahedron centroids in the same region as in
  Fig.~\ref{fig:particles}. The size of the dots is reduced in order
  to approximately compensate for the larger number of dots.}
\label{fig:centroids}
\end{figure}
\begin{figure*}
\includegraphics[width=16.cm]{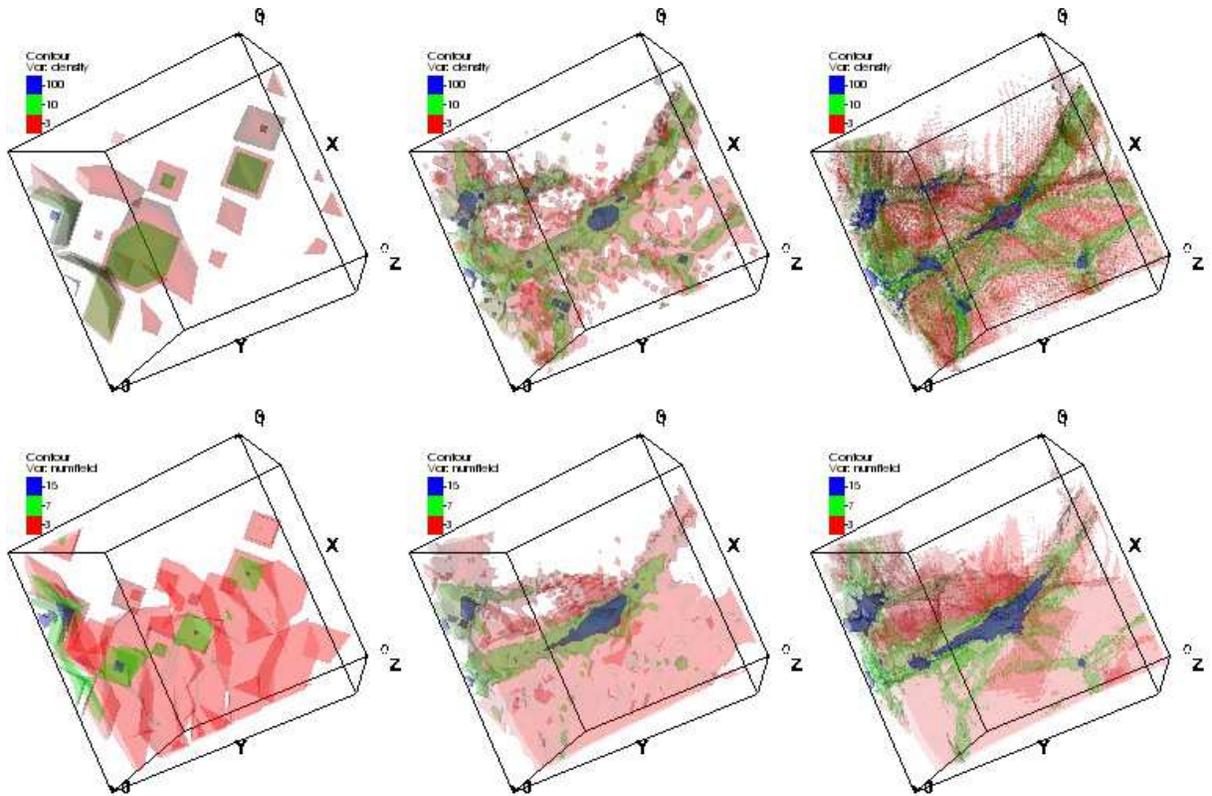}  
\caption{Same box as in Figs.~\ref{fig:particles} and \ref{fig:centroids}.
The top and bottom panels show the density and multi-stream fields
respectively. The fields in the left column are sampled on an $8^3$
grid, in the middle column on a $32^3$ grid, and in the right column
on a $128^3$ grid. The three density contours are: 3$\bar{\rho}$ --
red, 10$\bar{\rho}$ -- green, and 100$\bar{\rho}$ -- blue, and the
three multi-stream contours are: 3 -- red, 7 -- green, and
 15 -- blue. One can clearly see  how  filaments and pancakes emerge
 with increasing resolution from the left  to right. }
\label{fig:multi-stream-6p}
\end{figure*}
\begin{figure}
\includegraphics[width=7.cm]{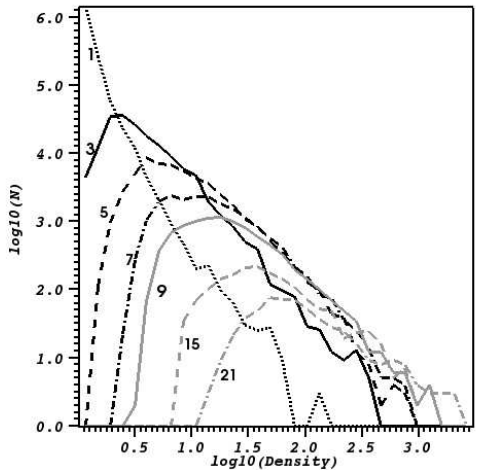}
\caption{Number of sampling mesh nodes as a function of density in the
  regions with number of streams shown next to each curve.}
\label{fig:hist-den-str-8-16}
\end{figure}
\begin{figure*}
\includegraphics[width=12cm]{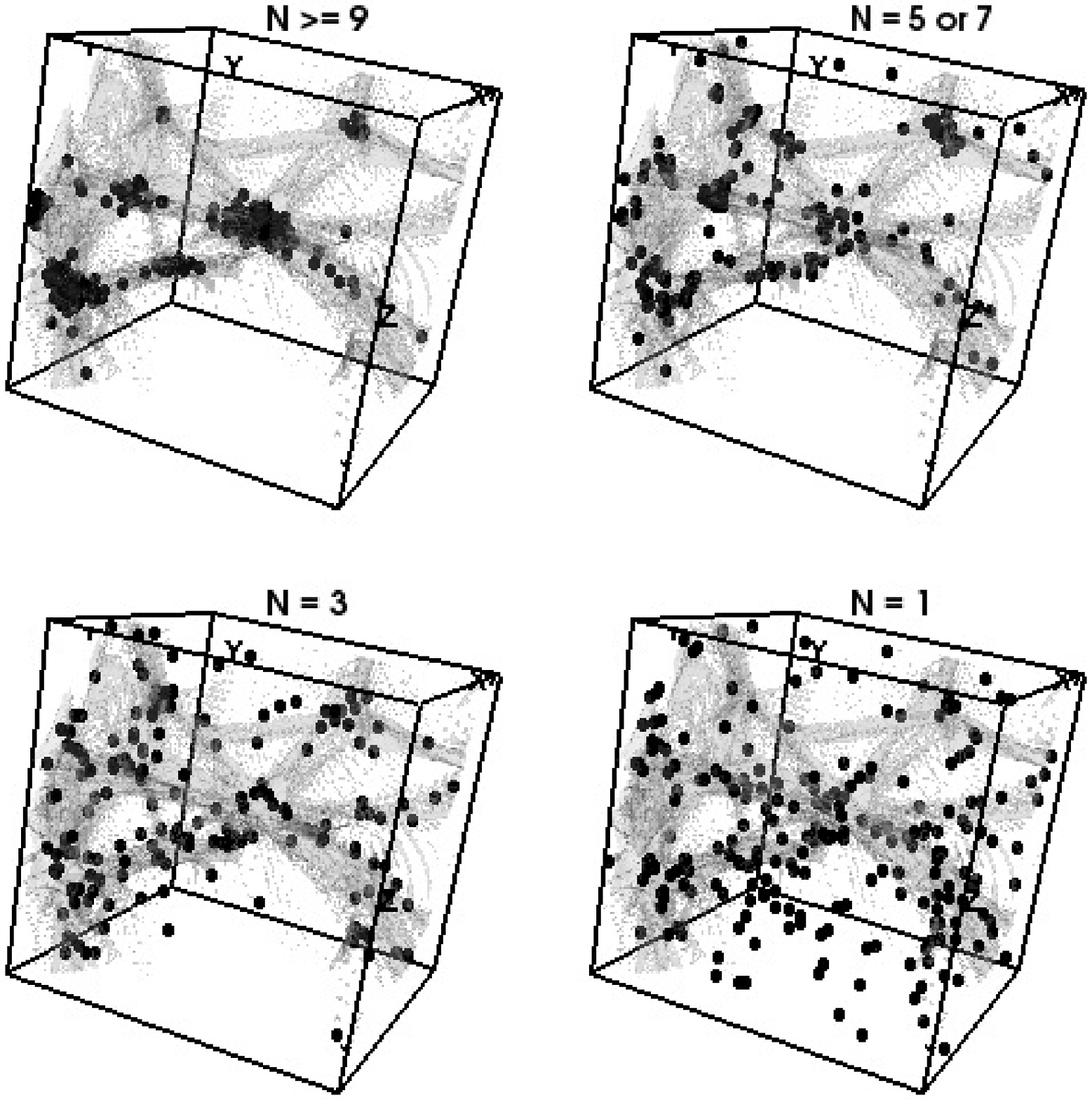}
\caption{Particles lying in regions with different numbers of streams
  vs density. The box is  the same as in the previous figures. All
  four panels show the density  contour $\rho/\bar{\rho} =10$.  
The panels show particles lying in $\ge$ 9-, 5- or 7-, 3-,  and
one-stream flows.} 
\label{fig:den-part-4p}
\end{figure*}

\subsubsection{Vertices vs. tetrahedra centroids}
We first compare the spatial distributions of particles (used as the
vertices of the tetrahedra) with the distribution of the tetrahedra centroids. 
The coordinates of the centroids are easy to compute since they are
simply the mean values of  the vertex coordinates: $ {\bf x}_c= 1/4
\sum_{i=1}^{4} {\bf x}_{v,i}$. If the density in a tetrahedron is
uniform then its centroid is also the center of mass of the
tetrahedron.  

Figures~\ref{fig:particles} and \ref{fig:centroids} show the spatial
distribution of the particles, i.e., vertices and centroids of the
tetrahedra in the chosen box. The number of centroids is five times
greater than that of the vertices since each elementary cube is
decomposed into five tetrahedra. The visual comparison of the figures
suggests that the distribution of centroids appears more filamentary
than that of the vertices. It also hints at the presence of a few very
thin sheets spanning the entire box. Unfortunately this can only be
fully appreciated when the distribution is manipulated (e.g. rotated)
with three-dimensional visualization software on a computer screen
\footnote{In this study, two visualization and plotting packages have
  been used: Mayavi (http://code.enthought.com/projects/mayavi/) and
  VisIt (https://wci.llnl.gov/codes/visit/).}.

\subsubsection{Density field vs multi-stream field}

We now show the multi-stream and density fields computed using the
tessellation method in  Fig. \ref{fig:multi-stream-6p} 
in the box depicted in Figs.~\ref{fig:particles}
and \ref{fig:centroids}. The overall parameters of the multi-stream field computed
on a 128$^3$ uniformly-sampled grid are as follows:  
$n_{\rm s,min} = 1$, $n_{\rm s,med} = 1$, $\langle n_{\rm s}\rangle =
1.45$, ${\rm std}(n_{\rm s}) = 1.93$,  $n_{\rm s,max} = 141$.
The overall parameters of  the density field within the selected box
(also computed on a 128$^3$ grid) are as follows:  
$\rho_{\rm min} = 0.032 \bar{\rho}$,
$\rho_{\rm med} = 0.16 \bar{\rho}$, $\langle\rho\rangle = 1.78
\bar{\rho}$, ${\rm std}(\rho) = 38.5 \bar{\rho}$,  
$\rho_{\rm max} = 2.2\times10^4 \bar{\rho}$,
where $\bar{\rho}$ is the mean density in the Universe and
$\langle\rho\rangle$ is the mean density within the chosen box.

Figure~\ref{fig:multi-stream-6p} demonstrates the crucial role of
spatial resolution.  The multi-stream and density fields computed with
increasing resolution on sampling meshes with $8^3$, $32^3$ and
$128^3$ nodes are shown in the bottom and top panels
respectively. It is remarkable that the three-stream flows clearly
seen as a pancake in the multi-stream field in the bottom left  panel 
at the lowest spatial resolution of 1$h^{-1}$Mpc
but not seen in the density field in top left panel.
The filaments and pancakes appear
in the middle column and acquire more details in the right column. This
allows us to tentatively conclude that the pancakes and filaments in
the box are thinner than $\sim 1h^{-1}$Mpc and some parts of the
structure are even thinner than $\sim 1/4{\rm h}^{-1}$Mpc.

As in the case of the multi-stream field, the density field clearly
reveals structure in more detail as resolution is increased. Note that
the multi-stream and density fields are not identical as stressed in
Ref.~\cite{sh-11}.  The correspondence of the peaks in the density
field to the clumps in the particle distribution in
Fig.~\ref{fig:particles} is much stronger than that of the multi-stream
field.

Figure~\ref{fig:hist-den-str-8-16} shows the distribution of densities
in 1, 3, 5, 7, 9, 15, and 21-stream flow regions computed on the high
resolution mesh with 128$^3$ nodes. The horizontal axis is the
logarithm of the density. The vertical axis is the logarithm of the
number of nodes in each of 32 equal logarithmic bins in the range of
densities from $\rho = 0.03\bar{\rho}$ to $\rho =
3,000\bar{\rho}$. The peak and the bulk of the distribution shift
monotonically to higher densities as the number of streams is
increased; this is mostly due to the absence of low density sites in
regions with high numbers of streams.

The dependence of the high end of the distribution on the number of
streams is considerably weaker than that of the low end. The high end
of the distribution appears to converge to an envelope with a simple
power law $N \propto \rho^{-1.45}$.  We will show later below that
although this kind of trend is qualitatively similar in a much larger
box (of size 384$h^{-1}$Mpc), the slope of the power law
envelope is significantly steeper. This is a clear indication of
large-scale bias.


 \subsubsection{High resolution density field vs particle distribution} 
 Here we discuss how the high resolution density field is related to
 the spatial distribution of particles. In
 Fig.~\ref{fig:den-part-4p} we present four panels showing the
 density field and particle distribution. All four panels display the
 same density contour $\rho/\bar{\rho} = 10$.  The particles in the
 panels are selected according to the number of streams at the
 position of each particle. The contour surface is semitransparent to
 allow us to see particles inside the contour. The top left panel
 shows only particles that reside in the regions with nine or greater
 number of streams; the total number of these particles is $n_{\ge 9}
 = 338$. All of them lie inside the density contour $\rho/\bar{\rho} =
 10$, and they are concentrated in the clumps located at the junctions
 of the filaments.  The top right panel shows particles in five- and
 seven-stream flows, totalling $n_5+n_7=93 + 77 = 170$
 particles. Practically all of these particles lie within the
 filaments except a few that are in pancakes.  The bottom left panel
 shows $n_3=185$ particles residing in the three-stream flows
 only. The chosen density level $\rho/\bar{\rho} = 10$ shows only
 hints of pancakes and the chosen particles tend to be in those
 regions.  Finally, the bottom right panel shows only particles
 ($n_1=222$) located in the regions with only one stream \ie in voids.
 A closer inspection shows that no single particle lies inside the
 chosen density contour. (Unfortunately the projection effect in the
 figure does not allow one to appreciate this to the full.) The figure
 and these numbers are in qualitative agreement with
 Fig.~\ref{fig:hist-den-str-8-16}: there is an obvious correlation
 between the multi-stream and density fields but there is not a
 one-to-one correspondence.  It is worth being reminded that this
 particular box has a mass that is almost 80\% greater than that
 expected on average.  Therefore, the mass fractions in clumps,
 fialaments, pancakes, and voids $f_{\rm m,c}:f_{\rm m,f}:f_{\rm
   m,p}:f_{\rm m,v}$ $\approx$ $n_{\ge
   9}:n_{7,5}:n_{3}:n_{1}$$\approx$ 37\%:18\%:20\%:24\% may not be
 necessarily typical. We stress that this estimate is based on a rough
 reading of Fig.~\ref{fig:hist-den-str-8-16}, more reliable estimates
 will require a morphological analysis that will be reported on in a
 separate paper.

\section{Statistics of tetrahedra parameters}
In order to evaluate multi-stream, density or other fields using the tessellation
one must make use of the tetrahedra that are principle elements of the tessellation.
Their control over many properties of the derived field is considerably more elaborate than 
the type and size of the elements of a regular grid.
Therefore it is worth  examining some of their parameters in more detail.
In the tessellation approach, the information stored in the particle
coordinates is  used to compute various parameters of the set of
tetrahedra that are generated in Lagrangian space. The tetrahedra are
arranged in a semi-regular mesh in Lagrangian space but after mapping
to Eulerian space form an irregular partly overlapping mesh (see
Fig.~\ref{fig:meshes} for an illustration).  All other fields, e.g.,
multi-stream and density fields, are derived from the parameters of
the tetrahedra and therefore they are also specified on an irregular
mesh.  We have already described how the multi-stream and density
fields can be computed on a regular rectangular mesh of arbitrary
sampling resolution, or on an arbitrary set of points.  Here we study
the statistics of the basic parameters of the tetrahedra. The complete
description of an arbitrary tetrahedron requires six parameters (\eg
the lengths of all edges), thus the full description of the shapes
requires a five-dimensional parametric space.  We select only four
parameters consisting of the tetrahedron volume and three others that
characterize the shape. These `basic parameters' will be defined and
discussed further below. The statistics below are obtained using the results
from the 384$h^{-1}$Mpc box.

\subsection{Tetrahedra volumes}

There are only two kinds of tetrahedra in the tessellation of
Lagrangian space: `small', with volume $1/6$ and `large', with volume
$1/3$\,$h^{-3}$\,Mpc$^3$ (see Fig. \ref{fig:cube}) for the
simulation at hand, where the initial inter-particle separation is $1
h^{-1}$Mpc. The large tetrahedron is regular with edges equal to
$\sqrt 2$$h^{-1}$Mpc. The smaller tetrahedron has three mutually
orthogonal faces which are isosceles right triangles with catheti
equal to 1$h^{-1}$Mpc and the fourth face is an equilateral
triangle with sides equal to $\sqrt 2$$h^{-1}$Mpc. As the system
evolves, the tetrahedra change their sizes and shapes.

At the nonlinear stage of the evolution, the volumes of most tetrahedra 
undergo many cycles of  collapsing to zero followed by expansion. The
volume of a tetrahedron can be easily computed from the coordinates of
its vertices either  by using Eq.~(\ref{eq:d0}) or the vector algebra
expression. If three edges of a tetrahedron that meet at one vertex
are represented by vectors $\bf a$, $\bf b$ and $\bf c$, then its
volume is given by the scalar triple product  
\be
V = {1 \over 6} ~{\bf a}\cdot {\bf b}\times{\bf c}.
\label{eq:tetr-vol}
\ee The sign of the volume obviously depends on the mutual orientation
of the vectors and can change in the course of evolution. The change
of the sign of the volume in Eq.~\ref{eq:tetr-vol} indicates the
formation or vanishing of caustic elements which are the common faces
of the neighboring tetrahedra having volumes with opposite signs. As
the volume changes continuously it must pass through zero at some
instant of time -- this can happen only when one vertex of a
tetrahedron crosses the face formed by its other three vertices so
that all four vertices of the tetrahedron instantaneously lie on a
plane.  Since this face also belongs to the neighboring tetrahedron
the vertex that crossed the face now lies in the interior of the
neighboring tetrahedron. Therefore the number of streams is increased
in the overlapping part by two, because the vertex is also shared by
several other tetrahedra that therefore also overlap with the
neighboring tetrahedron. The three-dimensional hypersurface can be
wrapped in an arbitrarily complex manner in six-dimensional phase
space but it must never be torn, nor ever intersect itself.

\begin{figure}
  \centering\includegraphics[width=8.cm]{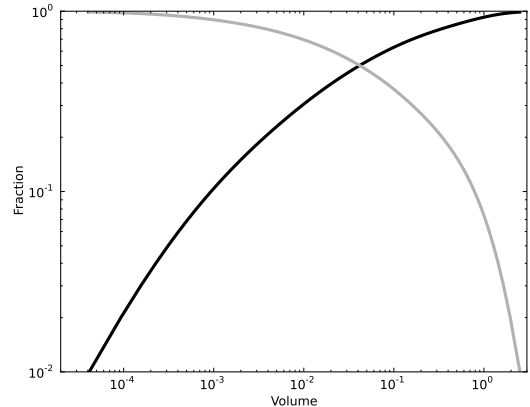} 
\caption{Bulk of the CDF of the volumes of the tetrahedra in the range
  from 1\% to 99\%.  The black and gray lines show the CDF, $F(V)$,
  and $1-F(V)$ respectively.} 
\label{fig:vol-bulk}
\end{figure}

The total Eulerian volume $V_E$ of all tetrahedra grows in the course
of evolution \be V_E = \sum V_i > V_L, \ee where $V_i$ is the Eulerian
volume of the $i$-th tetrahedron and the sum is taken over all
tetrahedra; $V_L$ is their total Lagrangian volume. This is not
unexpected since many tetrahedra overlap while filling the whole
comoving volume -- which is equal to the volume of Lagrangian space --
without gaps (as can be seen in the two-dimentional illustration in
Fig.~\ref{fig:meshes}).  In the simulation studied here the
three-dimensional manifold increased its volume by about 20\% compared
to its Lagrangian volume.

The cumulative distribution function (CDF) of the volumes of
tetrahedra, $F(V)$ is shown by the black line in
Fig.~\ref{fig:vol-bulk}, the gray line shows the behavior of $1 -
F(V)$. This combination allows us to display the behavior of the CDF
at small and large values simultaneously, the crossing point marking
the median value $V_{\rm med} = 0.042$$h^{-3}$Mpc$^3$ (see also
Table~\ref{tab:tetr-sizes} for other parameters). In the current
simulation, about 30\% of the tetrahedra have greater comoving volumes
than at the initial time.


\subsection{Sizes of tetrahedra}
We now turn to a simplified characterization of the shapes of
tetrahedra using only three size parameters that may be labeled as
`thickness', `breadth', and `length' as they satisfy the inequalities
$L > B > T$. We first select the face with the largest area as the
base of the tetrahedron.  The height perpendicular to the base is the
shortest of all heights, therefore we label it as the thickness $T$ of
the tetrahedron.  The shortest height of the base triangle is then
labeled as the breadth $B$, and finally, the longest edge of the base
triangle becomes the length $L$ of the tetrahedron.  The volume of the
tetrahedron can be expressed in terms of $L,B,T$ as $V=1/6 \,L\, B\,
T$.

\begin{figure}
  \centering\includegraphics[width=8.cm]{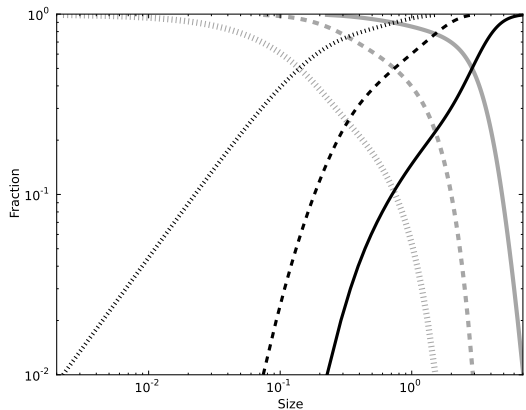}
\caption{CDFs F(T), F(B) and F(L)  are shown as the black dotted,
  dashed and solid lines respectively. The corresponding gray lines
  represent  $1-F$ for each size parameter. }
\label{fig:size-bulk}
\end{figure}

The cumulative distribution functions of the basic tetrahedra sizes --
the lengths, breadths, and thicknesses -- are shown in Fig.~\ref{fig:size-bulk}. 
The median, mean and standard deviation of these
parameters are given in Table 1 (the volumes are in units of
$h^{-3}$Mpc$^3$ and lengths in units of $h^{-1}$Mpc). Some tetrahedra
at the final stage have greater comoving sizes than at the initial
time: 17\%, 34\% and 80\% of tetrahedra have greater thickness, T, or
breadth, B, or length, L respectively.  Table \ref{tab:tetr-sizes}
provides also the range of the distribution between 1\% of the lowest
and and 1\% of the highest values.

\begin{table}          
  \caption{\label{tab:tetr-sizes} Statistical characteristics of basic parameters.}  
\begin{ruledtabular}

\begin{tabular}{| l | l | l | l |c|l|} 
   & Med. &   $<>$   &   std    &    1\% $\leftrightarrow$ 99\% & units \\ 
\hline
 Volume      &  0.042   &  0.25      &  0.52    & $5\times10^{-5} \leftrightarrow 2.0$ & $h^{-3}$Mpc$^{3}$\\
Thickness   &  0.14     &  0.27      &  0.33    & $2\times10^{-3} \leftrightarrow 1.4$ & $h^{-1}$Mpc\\
Breadth       &  0.74     & 0.94       &  0.72    & $8\times10^{-2} \leftrightarrow 2.8$ & $h^{-1}$Mpc\\
Length        &  2.88      &  2.88     & 1.58     & $2\times10^{-1} \leftrightarrow 6.9$ & $h^{-1}$Mpc \\
 \end{tabular}
\end{ruledtabular}
\end{table}

\subsection{Cross-correlations between the tetrahedra basic parameters}
The basic geometrical parameters of the tetrahedra characterizing
their sizes and shapes \ie the volume, thickness, breadth, and length
are obviously not statistically independent.  Table 1 shows the
cross-correlation coefficients ($C(P_i,P_j)$) between parameters $P_i$
and $P_j$, where $P_i = V, T, B, ~{\rm or}~ L$. Although their
distribution functions are not Gaussian as demonstrated by
Figs. ~\ref{fig:vol-bulk} and \ref{fig:size-bulk}, these numbers
provide a certain sense of statistical dependence between the
parameters. 
\begin{table}          
  \caption{\label{tab:cross-corr} Cross-correlations between basic parameters.}  
\begin{ruledtabular}
\begin{tabular}{|l|l|l|l|} 
                      & Thickness  & Breadth &  Length \\ 
\hline 
Volume         &  0.88           &  0.67     &  0.38 \\
\hline 
Thickness     &                    &  0.61      &  0.31 \\
\hline 
Breadth         &                    &              &  0.52\\
 \end{tabular}
\end{ruledtabular}
\end{table}

Despite the fact that the length, breadth, and thickness of a
tetrahedron enter the expression for the volume in a symmetric manner
their statistical roles are quite different, as Table 2 illustrates.
The volumes of tetrahedra are much more strongly statistically coupled
with the thicknesses ($C(V,T)= 0.88$) than with lengths
($C(V,L)=0.38$).  The coupling of the volumes with breadths is
stronger than with the lengths but weaker than with the thicknesses:
$C(V,L) < C(V,B) < C(V,T)$.  The strongest correlation among three
sizes of tetrahedra is observed between thickness and breadth and the
lowest between thickness and length: $C(T,L) < C(L,B ) < C(T,B)$.

\subsection{Low value limit of tetrahedra CDFs}

\begin{figure}
 \centering\includegraphics[width=8.cm]{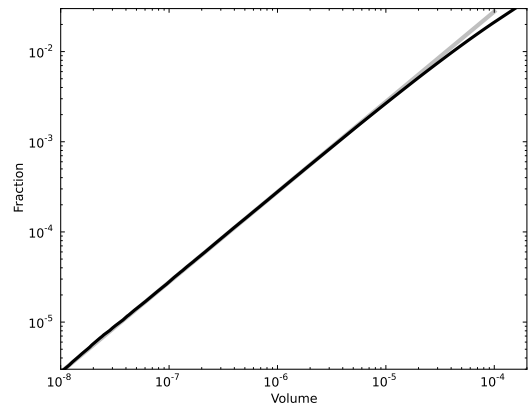} 
\caption{CDF, $F(V)$, of  3\% of the least voluminous tetrahedra.
The gray straight line is the power law $F \propto V$.  }
\label{fig:vol-low}
\end{figure}

\begin{figure}
  \includegraphics[width=8.cm]{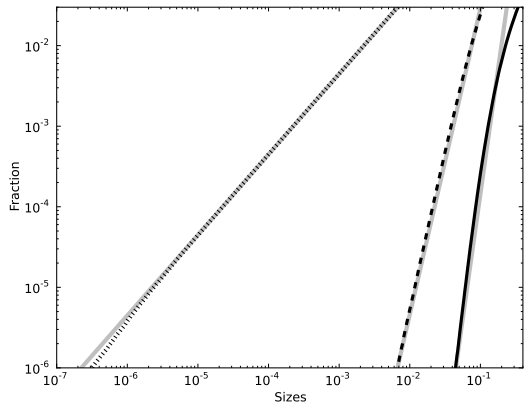}
\caption{CDFs, $F(T)$, $F(B)$, and $F(L)$ of the smallest 3\% of
  the corresponding values are shown by the dotted, dashed and solid
  black lines respectively. The gray solid straight lines show
  $F(T)\propto T$, $F(B)\propto  B^{3.8}$ and $F(L)\propto L^{6.2}$
 respectively.  }
\label{fig:size-low}
\end{figure}

The low value tails are shown in Figs.~\ref{fig:vol-low} and
\ref{fig:size-low}.  The gray straight lines are power laws
providing visual guidelines for the shapes of the curves. The slopes
of the straight lines are $n_{\rm V}=1, n_{\rm T}=1, n_{\rm B}= 3.8 $
and $n_{\rm L}=6.2$.  The power law approximation works well for the
thickness, breadth and volume; it is a little worse for the length but
still provides a reasonable description. This behavior shows that the
tetrahedra become more anisotropic as their volumes and sizes get
smaller.

\begin{figure}
\begin{minipage}[t]{.99\linewidth}
  \centering\includegraphics[width=8.cm]{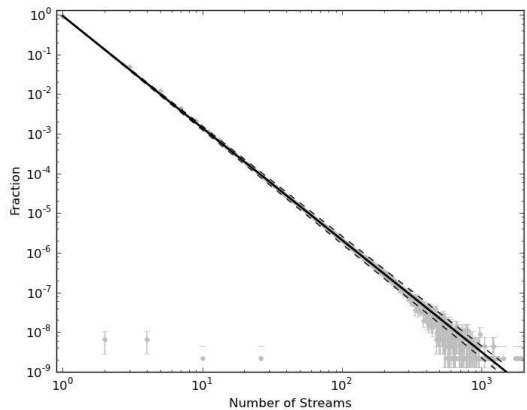}
\end{minipage}\hfill
\caption{Fraction of volume as a function of the number of
  streams. The data points with statistical error bars are shown in
  gray. The black solid and  two dashed lines show the fit  
$f_{\rm v}(n_{\rm s}) = 0.93 \, n_{\rm s} ^{-2.82\pm0.05}$. }
\label{fig:ns-384-2}
\end{figure}

\section{Multi-stream and density fields}
\subsection{Pdf of the multi-stream field}
As mentioned previously, at the lower end of the distance scale, the
multi-stream field is defined by the tessellation itself and, in
principle, can be computed at an arbitrary number of points, although
given a finite particle number, there will be an upper bound beyond
which no further useful information is obtained. At the other extreme,
one should make sure that the box is large enough to be statistically
representative. 

The results reported below are for a box size of $384^3{\rm
  h}^{-1}$Mpc with a sampling mesh of size $768^3$. (This corresponds
roughly to the dynamic range of the underlying N-body simulation.)
The overal statistical parameters of the multi-stream field, $n_{\rm
  s}$, as determined from this box, are as follows: ${\rm
  median}(n_{\rm s})=1.0, ~ \langle n_{\rm s}\rangle=1.27, ~{\rm std}(n_{\rm
  s})=2.2$.

Figure \ref{fig:ns-384-2} shows the fractions of volume in the regions
as a function of  the number of streams. Data points (with statistical
error bars) are shown in gray; the black solid line shows the fit
\be
f_{\rm v}(n_{\rm s}) = 0.93 \, n_{\rm s} ^{-2.82},
\label{eq:pdf-ns-fit}
\ee
and the dashed lines show the effect of changing the exponent by $\pm
0.05$.  The pdf has also been estimated by using another set of four
cubic boxes of different sizes: 32, 64, 128 and  256$h^{-1}$Mpc
with sampling meshes of variable size. The only difference found was
in the high value extent of the pdf:  from $n_{\rm s} \approx 30 $ for
the smallest sampling mesh  to $n_{\rm s} \approx 1000$ for the
largest number of sampling points (768$^3$), which is shown  in
Fig.~\ref{fig:ns-384-2}.  
 
Two features of the diagram are worth stressing.  First, the data
conform remarkably to the simple power law form of
Eq.~(\ref{eq:pdf-ns-fit}).  Second, since the number of streams can be
only odd (see \eg $\,$ Refs.~\cite{a-sh-z-82,sh-z-89}), the very small
volume fractions of the regions with even number of streams indicates
that the method is numerically robust and errors are negligible.  Four
spurious data points in the bottom left corner of the figure show the
fraction of the volume where the number of streams is even -- this
occurs in nine instances out of almost a half-billion sampling points.

The fractions of volume occupied by the most voluminous multi-stream
flows are given in Table \ref{tab:vol-mass-frac}. The single stream
regions provide a simple and robust method for defining
voids. Following this definition, we see that voids (defined up to the
resolution of the N-body simulation) occupy almost 93\% of the
simulation volume.  This number is somewhat greater than the estimate
of $\sim$86\% given in Ref.~\cite{ara-cal-etal-07} and is in
contradiction with the estimates of $\sim$17\% in
Ref.~\cite{hah-etal-07} and $\sim$13--82\% in Ref.~\cite{for-etal-09}.

While the definition of the fraction of volume occupied by voids is
unambiguously straightforward from the pdf of the multi-stream field
alone, the question of other principal morphological types
(pancakes/walls, filaments and clumps) is not as simple and requires
additional morphological analysis; this analysis will be carried out
elsewhere.

\begin{table}          
  \caption{\label{tab:vol-mass-frac} Fractions of volume and mass in streams with $n_s \le 15$.}  
\begin{ruledtabular}
\begin{tabular}{| l | l | l || l | l | l |} 
  n$_{\rm s}$  &  Volume               & Mass       &  n$_{\rm s}$  &  Volume                      & Mass   \\ 
\hline 
1         &  0.929                              &  0.242     & 3                   & 0.0484                         & 0.169   \\
\hline 
 5       &    0.0123                           &  0.108     &  7                  & 4.47$\times$10$^{-3}$ & 0.0714 \\
 \hline 
 9       &   2.09$\times$10$^{-3}$    & 0.0500   &  11               & 1.14$\times$10$^{-3}$  & 0.0374 \\
  \hline 
 13    &   6.96$\times$10$^{-4}$    &  0.0292    &  15              &  4.59$\times$10$^{-4}$  & 0.0234 \\
 \end{tabular}
\end{ruledtabular}
\end{table}

\subsection{Pdf of the density field}
The density field has been estimated from the tessellation in the
simulation box and with a  sampling mesh of the same size as for the
multi-stream analysis. The overall statistical parameters of the
density field are as follows:  
${\rm median}(\rho)=0.17\bar{\rho}, ~ \langle\rho\rangle=1.008\bar{\rho}, ~{\rm
  std}(\rho)=28.5\bar{\rho}$;  since the box in which the analysis is
carried out is somewhat smaller than the original simulation box, the
mean density is not exactly the mean density of the Universe,
$\bar{\rho}$. The cross-correlation coefficient of the density and
multi-stream fields is 0.45.

\begin{figure}
\includegraphics[width=8.5cm]{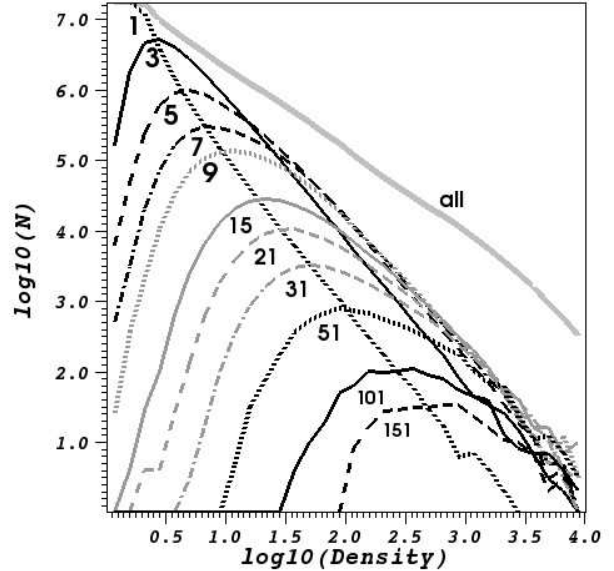}
\caption{Number of sampling mesh nodes as a function of density in
  regions with a fixed number of streams (values shown next to each
  curve).} 
\label{fig:hist-den-str-384-2}
\end{figure}

Figure \ref{fig:hist-den-str-384-2} shows the number of the sampling
mesh nodes, $N$, as a function of density computed at that node for a
number of different multi-stream flows in the range from one to 151
flows, as well as for the entire density field. The number of mesh
nodes, $N$, was computed for 32 equal logarithmic bins in the range of
densites from $\rho = 0.01\bar{\rho}$ to $\rho = 10,000\bar{\rho}$.

It was stressed in Ref.~\citep{sh-11} that although there is a
statistical correlation between the multi-stream and density fields,
no simple deterministic relation couples them together. This is
clearly demonstrated by the results in
Fig.~\ref{fig:hist-den-str-384-2}. As noted for the small box example
earlier, with the growth of the number of streams, the average density
is shifted to higher values mostly due to a considerable reduction of
the number of nodes with low densities. The high value tail of the
distribution functions again appears to converge to an approximately
limit that (very crudely) can be approximated by a power law $N
\propto \rho^{-1.8}$ which is significantly steeper than it was in the
small dense box. The distribution of densities also becomes flatter: the
full width at half maximum (FWHM) for the density distribution in the
101-stream flow is about 0.9 on the logarithmic scale, approximately
two times greater than that in the three-stream flow.

\begin{figure}
\includegraphics[width=9cm]{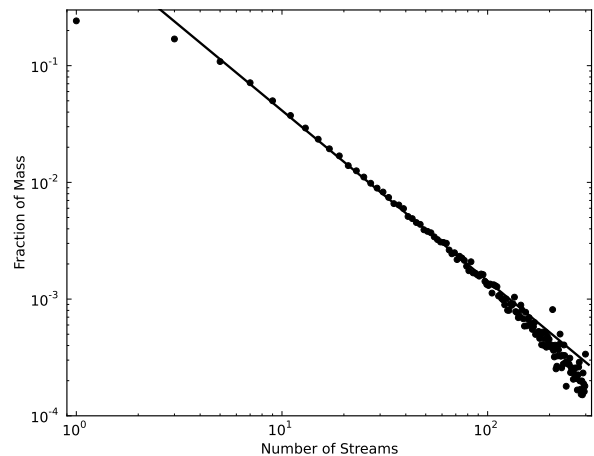}
\caption{Fraction of mass as a function of the number of streams
  is shown by the dots. The solid line is $F_{\rm m} = 1.19 \,n_{\rm
    s}^{-1.46}$.} 
\label{fig:mass-in-str-384-2}
\end{figure}
 
Figure \ref{fig:mass-in-str-384-2} shows the fraction of mass that
resides in the regions with a particular  number of streams (dots) and
the straight line is the power law 
\be
f_{\rm m} = 1.19 \,n_{\rm s}^{-1.46},
\ee
which provides a good fit to the data points in the range $5 \le
n_{\rm s} \lesssim 100$. Table \ref{tab:vol-mass-frac} provides the
list of fractions for several of the most massive multi-stream flows. 

The first points in Figs.~\ref{fig:ns-384-2} and \ref{fig:mass-in-str-384-2}
 (see also the top of Table~\ref{tab:vol-mass-frac})
corresponding to one-stream flow are easy to interpret: the voids
contain 24\% of mass and occupy 93\% of volume.  The flows with higher
number of streams do not correspond to a particular morphology,
although there is a general trend of the growth of average density
from pancakes to filaments and then to clumps or halos.

\section{Cosmological implications}
\subsection{Voids}
Identification of voids in the real Universe and in simulations is not
a simple problem for a number of reasons. Setting aside observational
issues such as sparse galaxy sampling, projection, and insufficent
survey volumes, even in simulations there is no universally accepted
notion of how to best define a void; a good example of the diversity
of current practice can be found in Ref.~\cite{colberg08}.

The multi-stream field provides a unique and simple physical
definition of voids as the regions with a one-stream flow (see
Ref.~\cite{sh-11} for additional discussion). This definition assumes
that voids are free from halos that can be resolved by the N-body
simulation, therefore it has an inherent small-scale cut-off,
interpretable as a cutoff in halo mass. As expected from hierarchical
structure formation, simulations with higher mass and force resolution
will find that the voids (of lower resolution simulations) are filled
with pancakes and filaments at smaller scales. (This can also be seen
in a simple way from the adhesion approximation~\cite{gur-sai-sh-89}.)

The particle mass (or the mass resolution) in the N-body simulation
used here is $ m_p \approx 3.6 \times 10^{10} M_{\odot}$, therefore
the one-stream flow region is free of halos with masses greater than
$\sim 10^{11}-10^{12}M_{\odot}$, since tens of particles are
sufficient to find a halo (though not to fully characterize it).  The
fraction of volume which is free of such halos is about 93\% and its
total mass fraction is about 24\% (Table
\ref{tab:vol-mass-frac}). Both numbers are in a good (although not
perfect) agreement with the corresponding estimates of
Ref.~\citep{ara-cal-etal-07} -- 86\% and 27\% respectively.

\subsection{Sheets/walls (pancakes) and other structures}
When the Zel'dovich approximation combined with catastrophe theory was
applied  to a collisionless medium of weakly  interacting particles, a
much richer picture than  that in the first paper by
Zel'dovich~\citep{z-70} emerged~\cite{a-sh-z-82,sh-z-89}. In this
picture, the pancakes were just the first multi-stream flows to be
formed in accordance with the gravitational instability. 
The pancakes correspond to a particular kind of singularity emerging
at the nonlinear stage; other singularities were associated with
filaments and compact clumps. This picture then found further
confirmation and development in the adhesion
approximation~\citep{gur-sai-sh-89,hid-10}.  

Ironically, pancakes were not observed in the first three-dimensional
simulation of the neutrino dominated universe~\citep{kly-sh-83}, nor
in countless simulations of CDM models of various flavors. Claims of
observing pancakes or sheets in cosmological N-body simulations have
emerged time after time in the literature, but most of them have been
anecdotal.  For instance, the authors of Ref.~\citep{hah-etal-07}
classified more than 45\% of volume as occupied by sheets (aka
pancakes), which strongly contradicts a visual impression. The authors
of Ref.~\citep{for-etal-09} attempted to improve the method used in
Ref.~\citep{hah-etal-07}; they found that the volume fraction occupied
by sheets ranged from 14\% to 60\% depending on the choice of
smoothing scale and other free parameters.  Even the lowest value is
difficult to reconcile with one's visual impression.  The geometrical
analysis based on the evaluation of partial Minkowski functionals
revealed that the dark matter density field in an \lcdm N-body
simulation showed a very weak signal of the presence of pancake-like
structures~\citep{sh-she-sah-04}. The result was essentially confirmed
by the analysis of the two largest superclusters in another N-body
simulation of the \lcdm cosmology~\citep{ara-cal-etal-10} analyzed by
a somewhat different method.

However, a new technique, the multiscale morphology filter (MMF)
~\citep{ara-cal-etal-07} based on the DTFE
methodology~\cite{sch-wey-00,sch-07} was able to unambiguously
identify walls in \lcdm N-body simulations.  The wall-like structures
were found to have the smallest fraction of mass $f_{\rm m,p}$=5.4\%
and second smallest fraction of volume $f_{\rm v,p}$=4.9\% after those
for clusters with $f_{\rm v,c}$=0.38\% .
 
If we believe the visual impressions from the study of the small box
in this paper then pancakes must be mostly associated with the
three-stream flows; this would imply a pancake volume fraction of
$f_{\rm v,p}$=4.8\% with $f_{\rm m,p}$=17\% (see Table
\ref{tab:vol-mass-frac}).  The former is in agreement with the
estimate in Ref.~\citep{ara-cal-etal-07} but the latter is more than
three times greater than their number.
 
In a similar fashion, assuming that the filaments are five- and
seven-stream flows, we can estimate their volume and mass fractions as
$f_{\rm v,f}$=1.6\% and $f_{\rm m,f}$=18\%, while the estimates of
Ref.~\citep{ara-cal-etal-07} are $f_{\rm v,f}$=8.8\% and $f_{\rm
  m,f}$=39.2\%. Again, there are clear discrepancies.  Finally, the
numbers for clumps are the remaining $f_{\rm v,c}$=0.7\% and $f_{\rm
  m,c}$=41\% by our estimates and $f_{\rm v,c}$=0.38\% and $f_{\rm
  m,c}$=28\% by Ref.~\citep{ara-cal-etal-07}.

We stress that there is an important difference between the pancakes
predicted by Zel'dovich and the structures usually called sheets and
walls in cosmological studies (see Fig. \ref{fig:multi-stream-6p} for an illustration ) . 
The Zel'dovich pancakes are defined as
three-stream flow regions -- as measured in this work -- while
structures called sheets and walls are always defined as a type of
density enhancement, and this was what was measured in
Ref.~\citep{ara-cal-etal-07}.  As we have already stated, there is a
positive -- but only modest -- correlation between density and
multi-stream fields, the cross correlation coefficient being 0.45, but
there is no one-to-one correspondence, as
Figs.~\ref{fig:multi-stream-6p}, \ref{fig:hist-den-str-8-16} and
\ref{fig:hist-den-str-384-2} demonstrate. This may well explain the
discrepancy, but only a more detailed direct comparison of the two
methods can finally resolve the issue.

Comparing the resulting mean overdensities for clumps, filaments,
walls/pancakes and voids we find 58.6, 11.2, 3.5 and 0.25
respectively, according to our estimates and 73, 4.45, 1.11 and 0.31
respectively, as found in Ref.~\citep{ara-cal-etal-07}. Thus, the
method based on using the multi-stream field systematically predicts
denser and more compact structures than the method based on the
density field, but less dense, and somewhat larger voids.

\subsection{Caustics}
Caustics in the dark matter density field have attracted much
attention in recent years because of the potential influence on the
(local) experimental detection of dark matter, see \eg
Refs.~\citep{moh-sh-06,moh-etal-07}. However, the study of
Ref.~\citep{vog-whi-11} has come to the conclusion that the number of
dark matter streams in the vicinity of the sun must be of the order of
10$^{14}$, which would completely negate the role of caustics.  If one
assumes this number, the predicted number of streams is absolutely
staggering: it means that if the galaxy began to form roughly
10$^{10}$ years ago, the average rate of creation of streams was
greater than one per hour. However, the number of the most massive
streams contributing half of the mass is 10$^{8}$ times less, \ie only
about a million streams in the vicinity of the sun.

Our new tessellation method allows us to identify caustic surfaces
directly.  The caustic surface is the boundary between neighboring
fluid elements, one of which has turned inside out one time more than
its neighbor. Therefore, if the volumes of the tetrahedra sharing a
common face are calculated using Eq.~(\ref{eq:d0}) or (\ref{eq:tetr-vol})
under the condition that the order of vertices be preserved as it was
at the initial stage, then the volumes of the tetrahedra separated by
a caustic would have opposite signs. The common face is the element of
a caustic surface. The tessellation not only provides an easy way of
finding caustic surfaces but also provides the triangulation of the
caustic for free. This allows us to compute the partial Minkowski
functionals for morphological analysis of the caustics as described in
Ref.~\citep{sh-she-sah-04,ara-cal-etal-10}.

Needless to say, the simulation used in this work was not designed for
counting numbers of streams in halos that host galaxies such as the
Milky Way, with a halo mass of $\sim 10^{12} M_{\odot}$. The mass
resolution of the simulation is $ m_p \approx 3.6 \times 10^{10}
M_{\odot}$ and our simulation has been carried out in a completely
different regime, where the number of streams reached only about a
thousand.  
 Of course one cannot expect to count 10$^{14}$ streams
directly simply because the number of particles in the simulations is
thousands times less than 10$^{14}$.  However, expecting counts of
streams up to a few million and greater in specially designed
simulations is quite reasonable.

\section{Discussion and Summary}
We have suggested a novel tessellation method for the analysis of
structure in the distribution of matter in cosmological N-body
simulations.  In this initial work, we focused on the relation between
density and multi-stream fields. The major features of the method are
as follows: 
\begin{itemize}
\item The method is physically motivated by the fact that the 3-D
  manifold in 6-D phase space describing the initial state of cold
  dark matter remains continuous under evolution driven by the
  gravitational instability.  
\item The analysis effectively uses the information that is stored in
  the coordinates of particles and their velocities \ie the entire
  phase space information. 
\item The underlying numerical code is simple and numerically robust.
\item The analysis does not involve any free parameters.
\item The density and multi-stream fields can be computed with high
  spatial resolution, limited only by the effective resolution of the
  underlying particle distribution. This technique can be easily
  adapted for other scalar, vector or other fields.
\item Any field can be computed on an arbitrary regular or point mesh.
\item The method allows finding  caustic surfaces directly regardless
  of their shapes. 
\item In addition, the method immediately provides a triangulation of
  caustic surfaces that allows for morphological analysis by the
  evaluation of the full set of partial Minkowski functionals.
\item To operate the code requires both initial and final
  coordinates. If the initial particles are the nodes of a regular
  mesh then they are obviously not needed. Following the evolution of
  the system in time is not required, the initial and final
  coordinates  are sufficient for all purposes.
\end{itemize}

Using the tessellation approach we demonstrated that the structure is
considerably more complicated than what can be seen directly in the
particle distribution, e.g., by comparing Fig.~\ref{fig:particles}
with Fig.~\ref{fig:multi-stream-6p}; the filaments and pancakes emerge
in both multi-stream and density fields as the resolution of the
sampling mesh is increased.

The multi-stream and density fields are modestly correlated (the
correlation coefficient is 0.45) which is qualitatively reflected in
Figs.~\ref{fig:hist-den-str-8-16} and
\ref{fig:hist-den-str-384-2}. Therefore, we confirm the conclusion of
Ref.~\citep{sh-11}, that the multi-stream field provides a
considerable amount of new information regarding cosmic structure.

The estimate of the fraction of volume and mass in voids, 93\% and
24\% respectivly, agrees reasonably well with the values of 86\% and
27\% obtained in Ref.~\cite{ara-cal-etal-07}.  The Zel'dolvich
pancakes are undoubtedly present among other generic structures \ie
filaments, clumps and voids in the \lcdm cosmology which is also in a
qualitative agreement with Ref.~\cite{ara-cal-etal-07}.  However, the
preliminary comparison of the volume and mass fractions of the other
components of the cosmic web -- \ie pancakes, filaments and clumps --
showed serious discrepancies.  This is likely due to differences in
the definition of (potentially heuristic) quantities such as pancakes,
filaments and clumps.  Our definitions are based on the multi-stream
field, while the definitions used in Ref.~\cite{ara-cal-etal-07} are
based on the density field.  In order to resolve this issue, further
morphological analysis is needed.  This work is underway and the
results will be reported separately.

To the best of our knowledge, the fractions of volume and mass in
multi-stream flows are estimated here for the first time. The pdf of
volumes demonstrates a remarkable power law scaling over the entire
range covering three orders of magnitude in $n_{\rm s}$ from one to a
thousand (Fig.~\ref{fig:ns-384-2}).  The pdf of masses can also be
approximated by a power law across most of its range
(Fig.~\ref{fig:mass-in-str-384-2}), though this scaling is not as well
fitted as the previous one.

In our approach, the evolution of the system is represented by the
evolution of the tetrahedra constructed by the tessellation of
Lagrangian space. The tetrahedra represent the major interface between
the particle coordinates and all other fields. Their control of the field
properties is considerably more elaborate than that of a regular mesh.
They also reveal the evolution of the volume elements of  dark matter
 in the gravitational clustering process.
Therefore it is
interesting to look at their parameters at the final time. We designed
three parameters that can be approximately considered as thickness,
breadth and length of a tetrahedron. The statistics of these
parameters are shown in Fig.~\ref{fig:size-bulk} and
\ref{fig:size-low} and clearly demonstrate a highly anisotropic
character of their deformations at the final time. The voids -- where
one might expect to find expanding tetrahedra -- contain about a
quarter of all mass (and therefore about a quarter of all tetrahedra),
however the breadths and lengths of the tetrahedra increased with
respect to the initial ones in 34\% and 80\% cases respectively, which
is significantly more than the fraction of the tetrahedra in voids.

The low ends of the CDFs of the tetrahedra volumes and sizes also look
like power laws, as shown in Fig.~\ref{fig:vol-low} and
\ref{fig:size-low}. CDFs of the volumes and thicknesses of the
tetrahedra are well approximated by a linear law which means that
their pdfs are flat at zero values.


\section{Acknowledgments}
SH and KH wish to acknowledge the use of supercomputing resources
under the Los Alamos National Laboratory Institutional Computing
Initiative. We also acknowledge support from the LDRD program at
Argonne National Laboratory.


\end{document}